\documentclass[prb, preprint, amsmath, amssymb, showpacs, superscriptaddress]{revtex4-1}
\usepackage[T1]{fontenc}
\usepackage{lmodern}
\usepackage{graphicx} % Include figure files
\usepackage{dcolumn}  % Align table columns on decimal point
\usepackage{bm}       % bold math
\usepackage{amsfonts}
\usepackage{dsfont}
\usepackage{csquotes}

\usepackage{ulem}
\usepackage{color}

\usepackage{physics}
\usepackage{siunitx}

\usepackage{hyperref}
\usepackage[dvipsnames]{xcolor} %for comment.
\renewcommand{\k}{\mathbf{k}}

\usepackage{xcolor}

\begin{document}
\title{Coherent control of chirality in Weyl semimetals}

\author{Amar Bharti}
\affiliation{%
Collective Dynamics and Quantum Transport Unit,
OIST Graduate University, Onna 904 0495, Japan}
\author{Margarita Khokhlova}
% \email[]{margarita.khokhlova@kcl.ac.uk}
\affiliation{Department of Physics, Kings College London, Strand, WC2R 2LS London, UK}
\author{Misha Ivanov}
\affiliation{%
	Max-Born Institut, Max-Born Stra{\ss}e 2A, 12489 Berlin, Germany }                     
\author{Gopal Dixit}
\email[]{gdixit@phy.iitb.ac.in}
\affiliation{%
Max-Born Institut, Max-Born Stra{\ss}e 2A, 12489 Berlin, Germany }
\affiliation{%
	Department of Physics, Indian Institute of Technology Bombay,
	Powai, Mumbai 400076, India }

\date{\today}


%%%%%%%%%%%%%%%%% END OF PREAMBLE %%%%%%%%%%%%%%%%
\begin{abstract}  
Weyl fermions in inversion-symmetric Weyl semimetals occur in pairs of opposite chirality, leading to symmetric optical responses under circularly-polarised light and a vanishing net photocurrent. 
Here, we show that tailored two-colour light fields break this symmetry and enable selective excitation of individual Weyl nodes. 
The interference between a circularly-polarised $\omega$ field and a phase-locked linearly-polarised $2\omega$ field generates a chirality-dependent redistribution of carriers in momentum space, resulting in a nonzero controllable photocurrent. 
We demonstrate that both the magnitude and sign of the photocurrent can be tuned via the relative phase and field strength of the two colours, and identify an optimal regime in which chiral selectivity is maximised. 
Our results establish a general route to optically-controlled chiral charge dynamics in Weyl semimetals using polarisation-structured light.
\end{abstract}

\maketitle 

\section{Introduction}
The chiral nature of Weyl nodes~\cite{Weyl1929, nielsen1983crystal} provides a route to using Weyl semimetals for optoelectronic and quantum information technologies~\cite{armitage2018weyl, Liu2020}. 
However, strategies for manipulating Weyl nodes, particularly for achieving node-selective excitation, remain limited~\cite{ma2017direct, chan2017photocurrents, yu2016determining, hosur2015tunable}.
Here, we demonstrate that a two-colour laser field with three-dimensional polarisation enables selective excitation of one of the two energy-degenerate Weyl nodes, depending on the node chirality.
This selectivity manifests as a residual photocurrent, i.e.,\ a DC current that persists after the end of the laser pulse~\cite{kruchinin2013theory}.

The response of Weyl nodes to circularly-polarised light~(CPL) at frequency $\omega$, with photon spin aligned along the axis connecting the nodes ($z$), is reminiscent of photoelectron circular dichroism~(PECD) in chiral molecules~\cite{Bowering2001, Lux2012, Powis2000, Fehre2021}. 
In gas-phase chiral molecules driven by CPL in the $xy$ plane, the emission of photoelectrons along the $z$ direction is asymmetric, generating a DC photocurrent~\cite{Bowering2001, Lux2012, Powis2000, Fehre2021}. 
This current has equal magnitude but opposite direction for the two enantiomers, providing a sensitive probe of molecular chirality~\cite{Lux2012, Fehre2021}. 
For an equal, racemic, mixture of enantiomers, the net current vanishes~\cite{Powis2000}.

A closely analogous effect arises in Weyl semimetals~\cite{chan2017photocurrents, ma2017direct, osterhoudt2019colossal, rees2020helicity, ni2021giant}. 
CPL at frequency $\omega$ induces local photocurrents at each Weyl node. 
Although the excitation rates are identical for the two nodes, the corresponding currents are equal in magnitude and opposite in direction, resulting in zero net photocurrent in the crystal~\cite{chan2017photocurrents, de2017quantized, golub2018circular}. 
Unlike CPL driving alone, phase-controlled two-colour fields break the symmetry between inversion-related Weyl nodes, enabling selective excitation of individual nodes.
We ask whether it is possible to lift this symmetry by introducing a control field linearly-polarised~(LP) along the $z$ axis, orthogonal to the polarisation plane $xy$ of the CPL driver.

\begin{figure}[!h]
\centering
\includegraphics[width=0.45\linewidth]{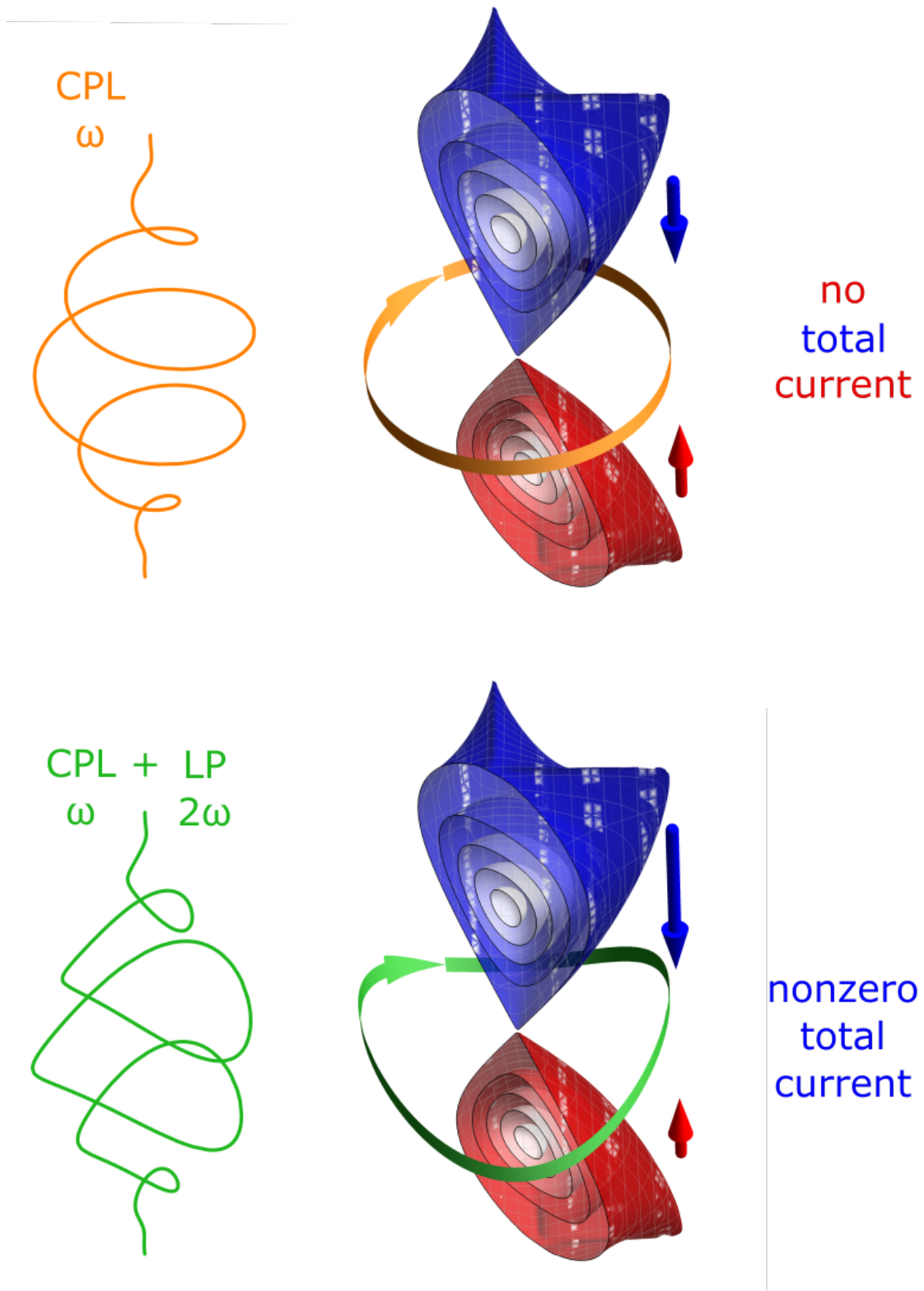
}
\caption{
{\bf {Schematics for the Weyl nodes driven by single- and two-colour field.}} 
A single-colour CPL field induces photocurrents equal in magnitude but opposite in direction, resulting in a zero net current. 
In a two-colour field, interference between CPL $\omega$ and LP $2\omega$ fields, controlled by the relative phase $\phi_{2\omega}$ between two fields, results in the asymmetric excitation of two Weyl nodes, resulting in a nonzero net photocurrent.}
\label{fig:fig0}
\end{figure}

Qualitatively, the photocurrent generated at each Weyl node can be understood as arising from an effective second-order (optical-rectification-like $\chi^{(2)}$-type) nonlinearity~\cite{morimoto2016topological, ma2019nonlinear}.
This naturally motivates the use of a control LP field at frequency $2\omega$, which has the double frequency of the CPL driver. 
By tuning the relative phase between the two fields, one can control the contributions of the $2\omega$ LP control and $\omega$ CP driving fields, which interfere constructively at one node and destructively at the other~\cite{mciver2020light}, see Fig.~\ref{fig:fig0}.
This interference enables the selective excitation of a single Weyl node and generates a net DC photocurrent.
Furthermore, control over the relative phase provides a mechanism for coherent manipulation of node selectivity. 
This approach is conceptually related to synthetic chiral light~\cite{ayuso2019synthetic}, developed for enantiomer-selective excitation in gas-phase molecules~\cite{Ayuso2021NatComm, Khokhlova2022, Mayer2024}. 
A key distinction, however, is that Weyl semimetals provide a crystalline environment with a fixed orientation, in contrast to the randomly-oriented molecular ensembles in the gas phase~\cite{Lux2012}.

\section{Results}
We begin by introducing the low-energy Hamiltonian describing linear dispersion near a Weyl node, 
$\mathcal{H} = \mathbf{v}\cdot(\mathbf{k}\bm{\sigma})$. 
Expanding around the two nodes of opposite chirality, we write
\begin{subequations}
	\begin{align}
		\label{eq:Eqlin1}
		\mathcal{H}_{+}(\mathbf{k}) & = - v_x k_{x}\sigma_{x} + v_y k_{y}\sigma_{y} - v_z \tilde{k}_{z} \sigma_{z}, \\
		\label{eq:Eqlin2}
		\mathcal{H}_{-}(\mathbf{k}) & = - v_x k_{x}\sigma_{x} + v_y k_{y}\sigma_{y} + v_z \tilde{k}_{z}\sigma_{z}. 
	\end{align}
\end{subequations}
Here $\tilde{k}_{z}$ denotes momentum measured relative to the corresponding Weyl node, $\mathbf{v}=(v_x,v_y,v_z)$ is the velocity vector, and $\bm{\sigma}$ is the vector of Pauli matrices. 
The Hamiltonians in Eqs.~\eqref{eq:Eqlin1} and~\eqref{eq:Eqlin2} describe Weyl nodes of chirality $\chi=1$ and $\chi=-1$, respectively, where the chirality is given by $\chi = \mathrm{sgn}(v_x v_y v_z)$. 
Such pairs of nodes with opposite chirality are a defining feature of Weyl semimetals~\cite{armitage2018weyl, lv2015experimental, morali2019fermi}.

For simplicity, we first consider isotropic dispersion, $v_x=v_y=v_z\equiv v$, corresponding to an untilted (type-I) Weyl semimetal. 
We further assume that both nodes lie at the Fermi level. 
In this case, chirality is the only distinguishing property of the two energetically degenerate nodes.
We first analyse the interaction of a Weyl node with CPL. 
The electron dynamics in the laser field are modelled using the semiconductor Bloch equations in the Houston basis~\cite{floss2018ab, krieger1986time, vampa2015semiclassical, bharti2022high, bharti2023weyl}, where the coupling between the Berry connection and the driving CPL governs the chiral response as discussed in Method section.

\begin{figure}[!h]
\centering
\includegraphics[width=0.6\linewidth]{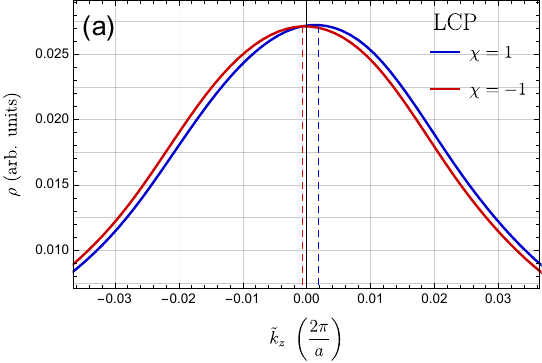}
\includegraphics[width=0.6\linewidth]{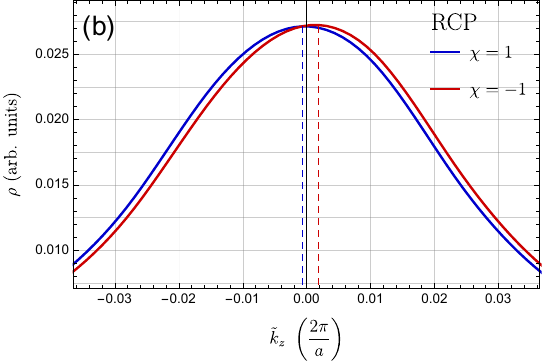}
\caption{
{\bf {Conduction band population and its asymmetry in single-colour field.}}
Residual electronic population in the conduction band $\rho$ induced by (a)~LCP and (b)~RCP laser pulse in the vicinity of the chiral Weyl nodes. 
For LCP, the residual population around the $\chi = 1$ node is skewed towards $+\tilde{k}_{z}$, whereas the population around the $\chi = -1$ node is slightly skewed towards $-\tilde{k}_{z}$, and vice versa for RCP. 
We use \SI{100}{fs} CPL pulse with intensity $\SI{e11}{W/cm^2}$ and wavelength $\omega=\SI{10.6}{\mu m}$ here and below if not specified otherwise.}
\label{fig:fig1}
\end{figure}

CPL selectively excites different regions of the node depending on its chirality~\cite{chan2017photocurrents, de2017quantized, rees2020helicity, bharti2023weyl}, see Fig.~\ref{fig:fig1}.
Left-CP~(LCP) light preferentially excites the $+\tilde{k}_{z}$ side of a node with $\chi=1$, as illustrated in Fig.~\ref{fig:fig1}(a), whereas right-CP~(RCP) light preferentially populates the $-\tilde{k}_{z}$ side of a node with $\chi=-1$, shown in Fig.~\ref{fig:fig1}(b).
The resulting carrier distributions at the two nodes are related by inversion symmetry, such that $\rho(-\tilde{k}_{z})$ at one node mirrors $\rho(\tilde{k}_{z})$ at the other.
Consequently, the photocurrents generated at the two nodes are equal in magnitude and opposite in direction~\cite{de2017quantized, golub2018circular}.
As a result, a Weyl semimetal with pairs of degenerate nodes produces no net photocurrent under CPL~\cite{chan2017photocurrents}. 
This cancellation motivates the search for schemes that break this symmetry and enable a nonzero current via chirality-selective excitation.

\begin{figure}[!h]
\centering
\includegraphics[width=0.6\linewidth]{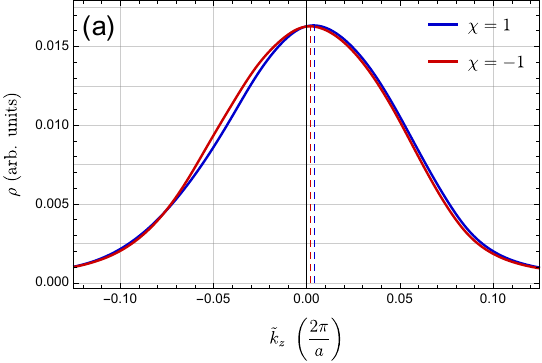}
\includegraphics[width=0.6\linewidth]{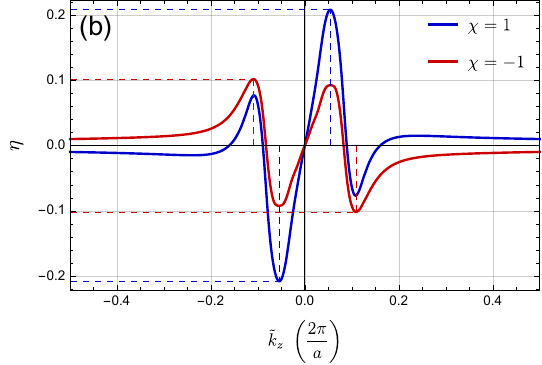}
\caption{
{\bf {Conduction band population and its asymmetry in two-colour field.}} 
(a)~Conduction band population $\rho$ around the chiral Weyl nodes for $\beta=0.2$. 
(b)~Asymmetry in the residual population centred around each Weyl node as a function of $\tilde{k}_{z}$.
The $\omega$ field is LCP.}
\label{fig:fig2}
\end{figure}

To lift this cancellation, we introduce a LP control field at frequency $2\omega$, oriented perpendicular to the CPL driver. 
The addition of a weak $2\omega$ field (with relative peak field strength $\beta=0.2$) significantly modifies the residual population, as shown in Fig.~\ref{fig:fig2}(a). 
In contrast to the single-colour CPL case, the populations at the two Weyl nodes are no longer related by mirror symmetry. 
In particular, the shifts of the population peaks away from $\tilde{k}_{z}=0$, previously observed for CPL, are strongly suppressed for the two-colour field~[Eq.~\ref{eq:pulse}], and the distributions are centred closer to $\tilde{k}_{z}=0$.

Moreover, the population becomes asymmetric with respect to $\tilde{k}_{z} \rightarrow -\tilde{k}_{z}$. For example, at the $\chi=1$ node, the population at $\tilde{k}_{z}=0.05$ exceeds that at $\tilde{k}_{z}=-0.05$, with a similar behaviour observed for the $\chi=-1$ node. 
Importantly, the asymmetry differs between the two nodes as reflected from Fig.~\ref{fig:fig2}(b), signalling a breakdown of chirality-related symmetry in the driven system.
To quantify this effect, we define an asymmetry parameter
\begin{equation}
\label{eq:asym}
    \eta = \frac{\rho(\mathbf{k}) - \rho(-\mathbf{k})}{\left[\rho(\mathbf{k}) + \rho(-\mathbf{k})\right]/2},
\end{equation}
where $\rho(\mathbf{k})$ and $\rho(-\mathbf{k})$ denote the residual populations induced by the combined CP $\omega$ and LP $2\omega$ fields for given LCP or RCP light and parameter $\beta$.

Figure~\ref{fig:fig2}(b) shows that even a weak $2\omega$ field induces a pronounced population asymmetry, which increases with $\beta$. 
Moreover, the asymmetry parameter $\eta$ exhibits a qualitatively different behaviour for the two Weyl nodes.
Examining $\eta$ on the $+\tilde{k}_{z}$ side reveals an asymmetric profile: for the $\chi=1$ node, $\eta$ displays a pronounced positive peak near $\tilde{k}_{z}=0$, followed by a shallow negative trough, resulting in an overall positive bias. 
In contrast, the behaviour for the $\chi=-1$ node differs markedly, indicating a strong dependence on chirality.
This imbalance implies a chirality-dependent redistribution of carriers~\cite{morimoto2016topological, ma2019nonlinear}, where $\rho(\mathbf{k})$ exceeds $\rho(-\mathbf{k})$ over a substantial region of the Brillouin zone.
As a consequence, the system supports a nonzero photocurrent~\cite{soifer2019band}. 
We therefore proceed to examine the dependence of this photocurrent on the control parameter $\beta$.

\begin{figure}[!h]
\centering
\includegraphics[width=0.7\linewidth]{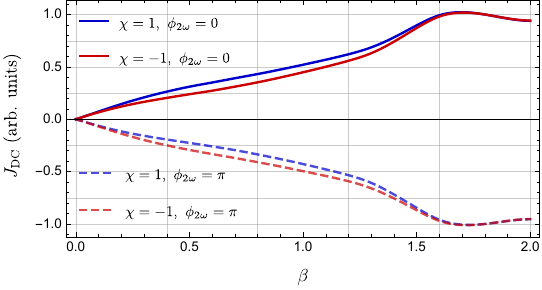}
\caption{
{\bf {Chirality of the photocurrent.}} 
Normalised photocurrent ${\mathrm{J}}_{\textrm{DC}}$ around both Weyl nodes as a function of the relative strength $\beta$ of two-colour laser fields for the relative phase $\phi_{2\omega} = 0$ (solid lines) and $\phi_{2\omega}= \pi$ (dashed lined). 
}
\label{fig:fig3}
\end{figure}

The chirality-dependent population asymmetry $\eta$ directly manifests as a photocurrent, shown in Fig.~\ref{fig:fig3}. 
The connection between the two can be expressed as 
\begin{equation}
\mathrm{J}_{\textrm{DC}} = \int_{\mathbf{k}} d\mathbf{k}\,[\rho(\mathbf{k}) - \rho(-\mathbf{k})] \, \pdv{\mathcal{E}(\mathbf{k})}{\mathbf{k}},
\end{equation}
where $\mathcal{E}(\mathbf{k})$ is the band dispersion~\cite{morimoto2016topological}. 
For $\beta=0$, corresponding to a single-colour CPL, the photocurrents generated at the two Weyl nodes cancel exactly, resulting in zero net current.
For small $\beta$, the contributions from the two nodes remain nearly indistinguishable, and the net photocurrent is negligible. 
As $\beta$ increases, however, the photocurrent becomes appreciable and develops a clear chirality dependence, with distinct contributions from each node. 
In this intermediate regime (e.g.,\ $ 0.2 \lesssim \beta \lesssim 1.6$), the photocurrent is strongly chirality-sensitive. 
For larger $\beta$, where the $2\omega$ field dominates, the photocurrent becomes progressively less sensitive to chirality. 

Nevertheless, over a broad range of $\beta$ the contributions from the two Weyl nodes remain chirally-sensitive, as shown in Fig.~\ref{fig:fig3}. 
This behaviour reflects the underlying asymmetry in the population: as seen in Fig.~\ref{fig:fig2}(b), the $\chi=1$ node exhibits a larger $\eta$, leading to a correspondingly stronger photocurrent. 
The photocurrent can be controlled by tuning the phase $\phi_{2\omega}$ between the $\omega$ and $2\omega$ fields. 
In Fig.~\ref{fig:fig3}, the sign of the photocurrent reverses as $\phi_{2\omega}$ changes from $0^\circ$ to $180^\circ$. 
For $\beta=0.5$, the $\chi=1$ node yields a photocurrent $\mathrm{J}_{\textrm{DC}} \approx 0.3$ at $\phi_{2\omega}=0^\circ$. 
Upon flipping the phase to $\phi_{2\omega}=180^\circ$, the photocurrent reverses sign, with the $\chi=-1$ node contributing an equal-magnitude current of opposite direction. 
These results demonstrate that the photocurrent is jointly controlled by the helicity of the combined $\omega-2\omega$ field and by the chirality of the Weyl nodes.

\begin{figure}[!h]
\centering
\includegraphics[width=0.6\linewidth]{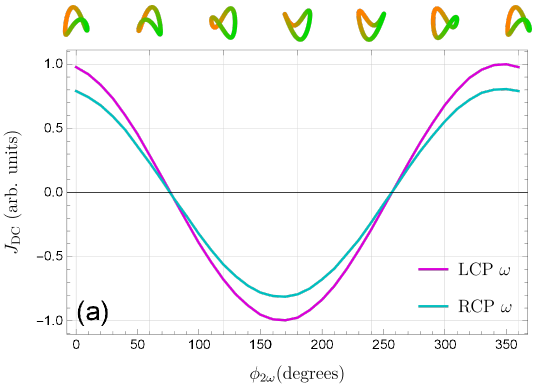}
\includegraphics[width=0.6\linewidth]{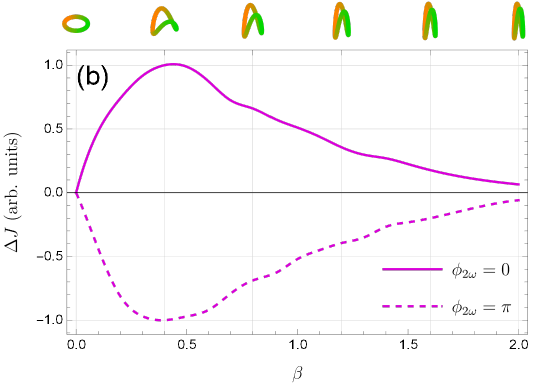}
\caption{
{\bf {Sensitivity of photocurrent with parameters of $\omega-2\omega$ field in a realistic Weyl semimetal.}} 
(a)~Variation in photocurrent along $\tilde{k}_{z}$ as a function of $\phi_{2\omega}$ for LCP and RCP $\omega$ field and $\beta = 0.5$.  
(b)~Sensitivity of the photocurrent asymmetry for LCP and RCP case as a function of the $2\omega$ field relative strength $\beta$. 
The Lissajous figures of the corresponding two-colour field which define the subcycle dynamics are shown on top of each figure.
Weyl semimetal is described by the full Hamiltonian, given in Eq.~\eqref{eq:ham_full}.}
\label{fig:fig4}
\end{figure}

So far, we have analysed the population dynamics and photocurrents induced by the $\omega-2\omega$ field within the linear Weyl Hamiltonian, Eq.~\eqref{eq:Eqlin1}--\eqref{eq:Eqlin2}. 
To assess the generality of our results, we consider an inversion-symmetric lattice model of a Weyl semimetal described by the Hamiltonian (see Method section). 
%\purple{In this model, the Weyl nodes are degenerate and located at the Fermi energy. (CAN BE REMOVED)}
We now examine the sensitivity of the net photocurrent with the two-colour field parameters in more detail and for the realistic Weyl Hamiltonian as given in Eq.~\eqref{eq:ham_full}. 
We first calculate the photocurrent as a function of $\phi_{2\omega}$. A pronounced photocurrent along the $\tilde{k}_{z}$ direction is observed for $\phi_{2\omega}=0^\circ$ with an LCP $\omega$ field, as shown in Fig.~\ref{fig:fig4}(a). 
The photocurrent decreases with an increase in $\phi_{2\omega}$, vanishing around $80^\circ$ and reaching a minimum near $170^\circ$, with a magnitude comparable to that at $\phi_{2\omega}=0^\circ$. 
The photocurrent reaches its maximum again close to $\phi_{2\omega}=350^\circ$.
Changing the helicity of the $\omega$ field from LCP to RCP significantly modifies the photocurrent, although the overall phase dependence remains similar.

To quantify this difference, we define a current asymmetry $\Delta_\mathrm{J} = \mathrm{J}_{\mathrm{LCP}} - \mathrm{J}_{\mathrm{RCP}}$, shown in Fig.~\ref{fig:fig4}(b) as a function of $\beta$. 
For $\phi_{2\omega}=0$, $\Delta_\mathrm{J}$ is positive, indicating a larger photocurrent for the LCP field. 
The asymmetry reaches a maximum near $\beta \approx 0.5$, demonstrating that the chirality-dependent photocurrent is most pronounced in this regime. 
Upon changing the phase to $\phi_{2\omega}=\pi$, the residual current $\Delta_\mathrm{J}$ reverses sign. 
For $\beta > 1$, $\Delta_\mathrm{J}$ approaches zero in both cases, indicating a loss of sensitivity to chirality as the $2\omega$ field dominates.
These results identify an optimal regime of intermediate field strengths, where chirality-selective photocurrent is maximised.

\section{Discussion}
We have demonstrated that two-colour light fields provide a route to selective excitation of Weyl nodes with opposite chirality.
While CPL alone generates equal and opposite photocurrents from the two nodes, resulting in complete cancellation, the addition of a phase-controlled $2\omega$ field breaks this symmetry and induces a nonzero chirality-dependent photocurrent.
We have shown that this effect arises from an asymmetric redistribution of carriers in momentum space, which can be tuned by both the relative strength and phase between the two colours. 
In particular, we identify an optimal regime of intermediate field relative strength, where the chirality-selective photocurrent is maximised, and we demonstrate coherent control of both the magnitude and sign of the photocurrent via the relative phase of the driving fields. 
These findings remain robust beyond the linear model Hamiltonian and persist for the tight-binding Hamiltonian. The phenomena remains robust for small tilt and energy split in Weyl nodes. 
Our results establish a general mechanism for controlling the chiral charge dynamics in Weyl semimetals using structured light. 
This approach opens new avenues for symmetry-resolved optoelectronics and for the selective manipulation of topological quasiparticles on ultrafast timescales~\cite{bao2021light, orenstein2021topology}.

\section{Methods}
To describe the interaction between the two-colour laser fields and the Weyl semimetal, we solve the quantum master equation in the Houston basis~\cite{mrudul2021light}:  
\begin{equation}
    i\pdv{}{t} \rho_{mn}^{\k}(t) = \left[\left(\epsilon_m^{\k_t} - \epsilon_n^{\k_t}\right)  -\frac{ \left(1 - \delta_{mn}\right)}{\mathrm{T}_2}\right]\rho_{mn}^{\k}(t) + \mathbf{E}(t) \cdot \left[\sum_l \left( \mathbf{d}_{ml}^{\k_t}~\rho_{ln}^{\k}(t) - \mathbf{d}_{ln}^{\k_t}~\rho_{ml}^{\k}(t) \right)   \right].
\end{equation}
Here, $\epsilon_{m}$ and $\epsilon_{n}$ denote the energies of electronic bands $m$ and $n$, respectively, while $\mathbf{d}_{mn}$ represents the transition dipole matrix elements between these bands. 
Both quantities are evaluated at the time-dependent crystal momentum $\k_t = \k + \mathbf{A} (t)$, , which defines the Houston basis, where $\k$ is the initial crystal momentum.
The total vector potential $\mathbf{A}(t)$ is related to the electric field $\mathbf{E}(t)$ via $\mathbf{E} (t) = -\partial{\mathbf{A}(t)}/\partial{t}$. 
 
The total vector potential, comprising a CPL driver field and a LP control field, is given by
\begin{equation}\label{eq:pulse}
    \mathbf{A}(t) = A_0 f(t)\left[ \cos(\omega t)\,\hat{\mathbf{e}}_x \pm \sin(\omega t)\,\hat{\mathbf{e}}_y + \beta \cos(2\omega t + \phi_{2\omega})\,\hat{\mathbf{e}}_z \right],
\end{equation}
where $A_0$  is the peak amplitude of the vector potential and $f(t)$ is a $\sin^2$ pulse envelope. 
The relative strength of the $2 \omega$ control field is scaled by the parameter $\beta$, and $\phi_{2\omega}$ dictates the subcycle phase difference between the CPL driver and the LP control field.
The signs $+$ and $-$ correspond to LCP and RCP light, respectively, and setting $\beta=0$ recovers the single-colour CPL case.

The initial state is taken to be a fully occupied valence band and an empty conduction band. 
The time-dependent coupled differential equations are then numerically propagated using a fourth-order Runge-Kutta method with a constant time step of \SI{0.8}{as}. 
Throughout this work, we use a phenomenological dephasing time $\mathrm{T}_2 = \SI{1.5}{fs}$, which is incorporated to account for electron-hole decoherence.
We sample the Brillouin zone using a $70 \times 70 \times 70$ $k$-point grid to perform numerical integrations. 
The electronic structure is described by a tight-binding model for an inversion-symmetric Weyl semimetal, given by the Hamiltonian
\begin{equation}
\mathcal{H}_{\textrm{Weyl}} = t\big[\cos(k_x a)\,\sigma_x + \cos(k_y a)\,\sigma_y + \{\cos(\tilde{k}_{z} a) - 0.5\,(1 + \sin(k_x a)\sin(k_y a))\}\,\sigma_z\big],
\label{eq:ham_full}
\end{equation}
where $t=\SI{1.8}{eV}$ is the hopping parameter and $a = \SI{6.28}{\AA~}$ is the lattice constant~\cite{bharti2023weyl}. The lattice vectors are $a_1 = (a,-a,0)$, $a_2 = (a,a,0)$ and $a_3 =(0,0,a)$.
Expanding the above Hamiltonian near the Weyl nodes yields a linearised, low-energy Hamiltonian given in Eqs.~(\ref{eq:Eqlin1}) and (\ref{eq:Eqlin2}).

\section{Acknowledgments}

M.K. acknowledges Royal Society research fellowship funding under URF\textbackslash R1\textbackslash 231460.

\bibliography{photocurrent.bib}
\end{document}